\def\cameraready{}
\def\cleanversion{}
\ifdefined\cameraready
  \documentclass[sigconf,screen]{acmart}
\else
  \input{review_class.tex}
\fi

\usepackage{microtype}
\usepackage{amsmath}
\usepackage[ruled,linesnumbered]{algorithm2e}
\usepackage{graphicx}
\usepackage{textcomp}
\usepackage{booktabs}
\usepackage{array}
\usepackage{subcaption}
\usepackage{tabularx}
\usepackage{xspace}
\usepackage{url}
\usepackage{xcolor}
\usepackage{listings}
\usepackage{multirow}
\usepackage{enumitem}
\usepackage[skins,breakable]{tcolorbox}
\usepackage{tikz}
\usepackage{pgfplots}
\pgfplotsset{compat=1.18}
\usetikzlibrary{arrows.meta, positioning, shapes.geometric, fit, backgrounds, calc}

\newcommand{\tool}{\textsc{SkillReuse}\xspace}
\newcommand{\bench}{\textsc{SkillReuse-Bench}\xspace}
\newcommand{\RQ}[1]{\textbf{RQ#1}}

\ifdefined\cleanversion
  \newcommand{\revision}[1]{#1}
\else
  \definecolor{revisionblue}{RGB}{0,64,160}
  \newcommand{\revision}[1]{\textcolor{revisionblue}{#1}}
\fi

\copyrightyear{2026}
\acmYear{2026}
\setcopyright{cc}
\setcctype{by-nc-nd}
\acmConference[ASE '26]{Proceedings of the 41st IEEE/ACM International Conference on Automated Software Engineering}{October 12--16, 2026}{Munich, Germany}
\acmBooktitle{Proceedings of the 41st IEEE/ACM International Conference on Automated Software Engineering (ASE '26), October 12--16, 2026, Munich, Germany}
\acmDOI{10.1145/3832783.3837468}
\acmISBN{979-8-4007-2882-2/2026/10}
\acmSubmissionID{ase26main-p820-p}
\received{2026-03-26}
\received[accepted]{2026-06-18}

\begin{document}

\title{Latent Reuse in Agent Skills: Multi-modal Clone Detection at Ecosystem Scale}

\ifdefined\cameraready
  \author{Jiaying Zhu}
  \orcid{0009-0007-4593-6207}
  \affiliation{%
    \institution{Nanyang Technological University}
    \city{Singapore}
    \country{Singapore}
  }
  \email{jiaying007@e.ntu.edu.sg}

  \author{Lyuye Zhang}
  \correspondingauthor
  \authornote{Corresponding author.}
  \orcid{0000-0003-3087-9645}
  \affiliation{%
    \department{College of Cryptology and Cyber Science}
    \institution{Nankai University}
    \city{Tianjin}
    \country{China}
  }
  \affiliation{%
    \institution{Nanyang Technological University}
    \city{Singapore}
    \country{Singapore}
  }
  \email{zh0004ye@e.ntu.edu.sg}

  \author{Wenbo Guo}
  \orcid{0000-0001-6655-8179}
  \affiliation{%
    \institution{Nanyang Technological University}
    \city{Singapore}
    \country{Singapore}
  }
  \email{honywenair@gmail.com}

  \author{Yang Liu}
  \orcid{0000-0001-7300-9215}
  \affiliation{%
    \institution{Nanyang Technological University}
    \city{Singapore}
    \country{Singapore}
  }
  \email{yangliu@ntu.edu.sg}
\else
  \input{authors_anonymous.tex}
\fi

\begin{abstract}
\revision{An agent skill is a reusable package for extending an LLM agent, typically a \texttt{SKILL.md} file that combines YAML metadata, natural-language instructions, and executable code.
Public repositories now host over two million skills, yet existing tools analyze each artifact in isolation, and registries do not track reuse created through copying, renaming, or adaptation.
Detecting these links is difficult because reuse may appear in one channel while the others change; conventional single-channel clone detectors can therefore miss such adaptations.
We present \tool, a multi-modal clone
detector that combines global lexical matching with channel-specific
representations for YAML, prose, and code, then uses logistic regression to
produce clone scores and interpretable clone-type labels. We also introduce
\bench, an annotated benchmark of 300 skill pairs spanning exact copies,
 renamed copies, adaptations, and semantic equivalents. On \bench, \tool
reaches an F1 of 0.939 with 0.952 precision, improving over TF-IDF and delivering 4.2$\times$ higher recall on Type-4 semantic
clones than MinHash. Applied to all 137,470 skills that pass the content filter, \tool identifies 1.06 million clone pairs involving 66.8\% of the analyzed skills; 95.3\% of these pairs cross author boundaries. Among skills in the analyzed name-based clone families, 67\% are superseded by a higher-quality variant.
Tracing 938 security-relevant skills through the clone graph surfaces
16,587 clone links spanning 6,376 related skills that per-skill scanners
alone would miss.}
\end{abstract}



\maketitle

\section{Introduction}
\label{sec:introduction}

\revision{An agent skill is a portable, modular package of instructions and
resources that gives an LLM agent a reusable capability, typically stored as a
\texttt{SKILL.md} artifact combining structured YAML frontmatter, free-form
natural-language guidance, and executable code snippets.} A developer can drop
such a file into an agent workspace and immediately add support for tasks
ranging from code review and reverse engineering to deployment automation.
Since public skill registries emerged in late 2025, the ecosystem has scaled
with unusual speed: more than \revision{2 million} skills are already available across
registries and GitHub repositories. Recent work has started to document this
growth and the broader structure of the ecosystem~\cite{skillsurvey,skilldataanalysis}.

What the ecosystem still lacks is basic support for tracking \emph{reuse}.
Skills usually omit dependency declarations, version histories, or provenance
fields that would reveal when one artifact was copied from another. As a
result, skill families form through invisible edits: authors duplicate a file,
rename it, translate the instructions, replace a code block, and publish the
result as a new skill. Without explicit lineage, maintainers cannot tell which
skills share the same underlying content, how much of the ecosystem is
redundant, or where a defective pattern has spread. \revision{The January 2026
ClawHavoc incident, which introduced more than 1,200 malicious skills into
OpenClaw~\cite{clawhavoc}, underscores the practical need for ecosystem
visibility: once suspicious content is identified, maintainers need a way to
find related artifacts across authors and registries.}

\revision{To our knowledge, no existing tool is designed to detect clones among agent skills. Current agent-skill analyzers inspect individual artifacts for security, structural quality, or malicious behavior~\cite{skillscan,skillfortify,maltool,skillject,skillinject,skillpromptinjection}, but do not recover relationships among skills. Traditional clone detectors primarily target homogeneous source-code artifacts~\cite{roysurvey,kamiyaccfinder,jiangdeckard,sajnanisourcerercc,zhaodeepsim}, whereas a \texttt{SKILL.md} file interleaves YAML metadata, natural-language instructions, and multi-language code, with evidence of reuse potentially confined to one channel. Our primary goal is therefore to develop and evaluate a clone detector tailored to these heterogeneous skill artifacts.}

\revision{We present \tool{}, a multi-modal approach that represents YAML, prose, and code separately, combines these channel-specific signals with a global lexical view and cross-channel interactions, and uses logistic regression to produce pairwise clone scores. The detected relationships also support two downstream analyses: registry inflation, which measures how much listed content is redundant, and, as a pilot application, security reach, which identifies related skills that may warrant inspection after a security-relevant pattern is found. The practical difference is substantial at both evaluation scales: on \bench{}, \tool{} detects 139 of 150 positive pairs, compared with 130 for flat TF-IDF and 101 for MinHash; over the 137,470-skill analysis corpus, exact hashing exposes 20,016 redundant skill entries (1.17$\times$ inflation), whereas \tool{} exposes 81,427 and reduces the corpus to 56,043 concepts (2.45$\times$ inflation).}

\revision{Hidden reuse is therefore a first-order property of the ecosystem. From the detected relationships, we construct
an ecosystem-level clone graph and use it for two concrete analyses: registry
inflation, which measures how much listed content is redundant, and
security reach, which identifies related skills that may warrant
inspection after a security-relevant pattern is found.
Across all 137,470 skills in the analysis corpus, \tool{} identifies 1.06 million clone pairs involving 66.8\% of the analyzed skills; 95\% of these links cross author boundaries, and the graph yields a conservative 2.45$\times$ registry-inflation estimate (Section~\ref{sec:ecosystem}). In the security application, starting
from 938 skills whose content matches standard security-relevant code
signatures, the clone graph surfaces 6,376 related skills, including 2,676 that
also match the seed's security-relevant pattern. Together, these results show what per-skill
inspection misses: the reuse structure connecting artifacts across the
ecosystem.}

This paper makes four contributions:
\begin{itemize}[leftmargin=*]
 \item \revision{To our knowledge, we provide the first ecosystem-scale study of undeclared reuse in agent skills, quantifying clone prevalence, cross-author relationships, and registry inflation over the full skill analysis corpus.}
 \item \revision{We introduce \tool{}, the first multi-modal clone-detection
 approach tailored to the heterogeneous structure of agent skills.}
 \item \revision{We show how a skill clone graph can, at ecosystem scale,
 extend a set of security-relevant matches to related artifacts that may warrant
 further inspection.}
 \item \revision{We release \bench{}, to our knowledge the first labeled
 benchmark for skill-clone detection: a 300-pair primary benchmark and a
 450-pair scale-robustness expansion, both stratified by difficulty.}
\end{itemize}

\section{Background and Motivation}
\label{sec:background}

\subsection{Agent Skills}

\revision{An agent skill is a reusable package that extends an LLM agent at runtime, most commonly through a \texttt{SKILL.md} artifact~\cite{skillsurvey,skilldataanalysis}. Its three in-file channels are YAML frontmatter for names, descriptions, category tags, compatibility constraints, and configuration; an NL body for activation, interaction, and behavioral instructions; and embedded, potentially multi-language code for operations such as shell commands, API calls, and file manipulation. This structure appears in over 95\% of skills across major registries; 78\% contain code, and virtually all contain NL instructions~\cite{skillsurvey,skilldataanalysis}. We therefore model YAML and code as optional channels but restrict the analysis to the focal \texttt{SKILL.md}, excluding external assets, linked files, and repository history.}

\revision{\noindent\textbf{Ecosystem Scale.} Skills circulate through marketplaces such as SkillsMP and repositories on GitHub without package management, versioning, or dependency declarations. Our SkillsMP snapshot contains over 2 million listings across 15 major categories, including code generation, testing, DevOps, security analysis, and documentation, and has grown 4$\times$ in six months. Reuse therefore occurs largely through copying and modifying \texttt{SKILL.md} files, creating undeclared clone relationships that registries do not track.}

\subsection{Clone Detection: From Code to Skills}

Code clone detection commonly distinguishes four clone
types~\cite{roysurvey}. Table~\ref{tab:clonetypes} presents these types
alongside our adapted definitions for multi-modal agent skills, which must
consider similarity across three content channels (YAML, NL, Code) rather
than a single source language.

\begin{table}[t]
\centering
\caption{Adapting the classical clone taxonomy to multi-modal skills.}
\label{tab:clonetypes}
\begin{tabularx}{\columnwidth}{@{}c>{\raggedright\arraybackslash}p{2.35cm}>{\raggedright\arraybackslash}X@{}}
\toprule
\textbf{Type} & \textbf{Code Clone} & \textbf{Skill Clone} \\
\midrule
\textbf{I} & Exact except layout/comments &
  Near-verbatim across YAML, NL, and code \\[2pt]
\textbf{II} & Renaming/literal edits &
  Cosmetic relabeling in metadata or code; structure preserved \\[2pt]
\textbf{III} & Statement-level edits &
  Partial edits in one or more channels; core workflow retained \\[2pt]
\textbf{IV} & Same semantics, different syntax &
  Capability-preserving transformed variant with cross-channel correspondence \\
\bottomrule
\end{tabularx}
\end{table}

\revision{For Type I, Anthropic's \texttt{docx} skill~\cite{skilldocxanthropic}
and davila7's fork~\cite{skilldocxdavila7} contain the same 20{,}056-character
\texttt{SKILL.md} across all three channels. The abridged diffs below illustrate
Types II--IV; unchanged lines are omitted.}
\begin{tcolorbox}[breakable,colback=gray!4,colframe=gray!45,boxrule=0.35pt,arc=1pt,
left=3pt,right=3pt,top=3pt,bottom=3pt]
\revision{
\textbf{(1) Type II---cosmetic relabeling.}\\[2pt]
Metadata (\texttt{slack-gif-creator}, ComposioHQ~\cite{skillslackgifcomposio} $\rightarrow$ davepoon~\cite{skillslackgifdavepoon}):\\[0pt]
\texttt{\ \ name: slack-gif-creator}\\[0pt]
\texttt{\ \ description: Toolkit for creating animated GIFs\ldots}\\[0pt]
\texttt{\textbf{+ category: creative-collaboration}}\\[1pt]
Only this line changes in the 17{,}056-character file.\\[5pt]
Code (synthetic \texttt{angular} renaming; Section~\ref{sec:benchmark}):\\[0pt]
\texttt{\ \textbf{-} import \{signal, ...\} from "@angular/core";}\\[0pt]
\texttt{\ \textbf{+} import \{signal, ...\} from "@angular-v2/core";}\\[1pt]
Only the module path is renamed.\\[6pt]
\textbf{(2) Type III---content extension:} \texttt{xlsx}
(ComposioHQ~\cite{skillxlsxcomposio} $\rightarrow$ danielmiessler~\cite{skillxlsxdaniel})\\[2pt]
\texttt{\textbf{+ \#\# Load Full PAI Context}}\\[0pt]
\texttt{\textbf{+ read \textasciitilde/.claude/skills/PAI/SKILL.md}}\\[2pt]
The fork adds $\approx$2,700 characters; the original 10,600-character workflow is unchanged.\\[6pt]
\textbf{(3) Type IV---cross-channel restructuring:} \texttt{idor-testing}
(zebbern~\cite{skillidorzebbern} $\rightarrow$ Dokhacgiakhoa~\cite{skillidordok}; Fig.~\ref{fig:example})\\[2pt]
\texttt{\ \ metadata.author: zebbern} preserved; description carried over\\[0pt]
\texttt{\ \ code blocks: 26} $\rightarrow$ \texttt{0};\quad NL body $\rightarrow$ 24 sub-skill links
}
\end{tcolorbox}
\revision{The fork examples have explicit GitHub provenance; the Type-II code
example is generated by the benchmark's renaming operator. All displayed lines
and channel counts come from the paired artifacts (Section~\ref{sec:benchmark}).}

\revision{Skill types are assigned from the joint YAML, NL, and code similarity
profile (Section~\ref{subsec:classification}), because a pair may exhibit a
different degree of similarity in each channel. The taxonomy describes
structural evidence rather than causal provenance. Type IV remains the least
certain category and excludes pairs linked only by topic or registry category.}

\revision{Existing detectors primarily target homogeneous source code: line- and token-based systems such as NiCad~\cite{roynicad}, SourcererCC~\cite{sajnanisourcerercc}, and CCFinder~\cite{kamiyaccfinder} emphasize surface similarity, while tree- and graph-based techniques~\cite{baxterclone,jiangdeckard,krinkeidentifying} require parseable AST or PDG input. Transformer models~\cite{fengcodebert,wangcodet} improve semantic-clone detection on benchmarks such as BigCloneBench~\cite{svajlenkobigclonebench}, but still assume source-code artifacts. They therefore do not directly capture the examples above, where clone evidence may appear in YAML or prose, or code may be moved into linked sub-skills.}

\subsection{Motivating Example}
\label{subsec:example}

\begin{figure}[t]
\centering
\includegraphics[width=1\columnwidth]{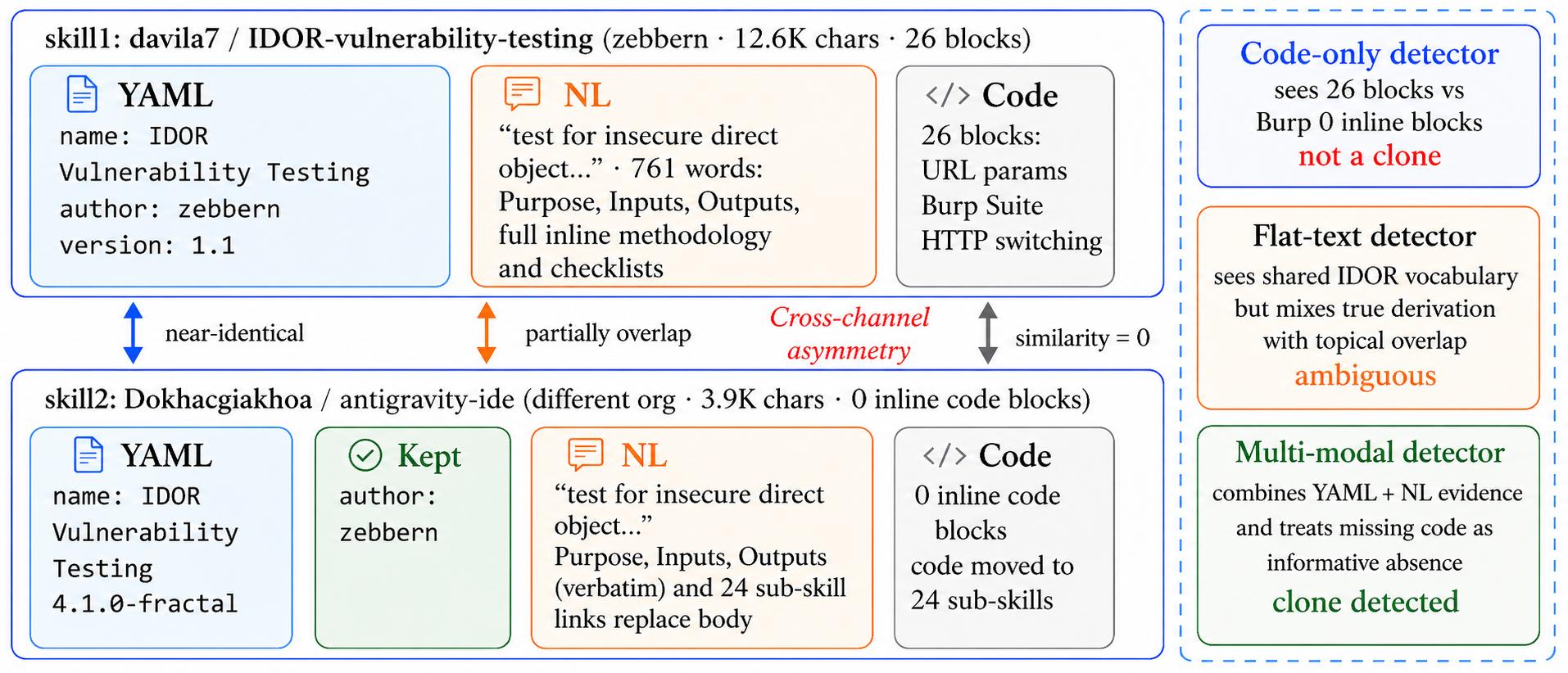}
\caption{A real clone pair exhibiting cross-channel asymmetry: YAML
metadata is near-identical, NL instructions partially overlap, but code similarity is zero (all blocks extracted to
sub-files). \revision{The explicit fork provenance confirms derivation.}}
\Description{Two related IDOR-testing skills are compared across YAML, natural-language, and code channels. Their YAML is nearly identical and their prose partially overlaps, but the second skill replaces 26 inline code blocks with links to 24 sub-skills. A code-only detector misses the relationship, whereas the multi-modal detector combines the remaining channel evidence.}
\label{fig:example}
\end{figure}

Figure~\ref{fig:example} shows a real clone pair discovered by \tool.
The original skill ($s_i$), a 12.6K-character IDOR vulnerability testing
methodology by \texttt{zebbern}~\cite{skillidorzebbern}, contains 26 code
blocks (51\% code by
bytes). A second author (\texttt{Dokhacgiakhoa}~\cite{skillidordok}) adapted this skill into a
modular format: the code blocks were extracted to 24 sub-skill files and the
NL body was replaced with links. Yet the \texttt{author:\ zebbern}
attribution and the trigger description were preserved verbatim---confirming
derivation despite the restructuring.

This pair exhibits \emph{cross-channel asymmetry}: YAML similarity is high
(${\sim}$0.95), NL is moderate (${\sim}$0.60), and code is zero. A code-only
detector like SourcererCC~\cite{sajnanisourcerercc} sees no code overlap and
misses the clone entirely. A flat-text detector conflates the similarity
with same-category vocabulary overlap. Only a multi-modal approach that
reasons about per-channel evidence---and treats the \emph{absence} of a
channel as a signal rather than noise---can reliably detect this type of
adaptation. \revision{Any shared flaw in the exploitation methodology could
therefore remain in the linked sub-skills while escaping code-only comparison.}

\section{Approach}
\label{sec:approach}

This section presents \tool, our approach for detecting multi-modal clone
relationships among agent skills. We first formalize the problem
(Section~\ref{subsec:problem}), give an overview of the approach
(Section~\ref{subsec:overview}), then detail multi-modal decomposition
(Section~\ref{subsec:decomposition}), modality-specific encoding
(Section~\ref{subsec:encoding}), cross-modal fusion
(Section~\ref{subsec:fusion}), and clone type classification
(Section~\ref{subsec:classification}).

\subsection{Problem Definition}
\label{subsec:problem}

\noindent\textbf{Input.} A corpus of $N$ agent skill documents $\mathcal{S} =
\{s_1, \ldots, s_N\}$, where each skill $s_i$ is a structured document
containing YAML metadata $y_i$, natural language instructions $n_i$, and
embedded code blocks $c_i = \{b_1, \ldots, b_k\}$.

\noindent\textbf{Output.} A set $\mathcal{C}$ of annotated clone pairs. Each
element $(s_i,s_j,\tau,\sigma)$ contains two skills, a type $\tau$ from Type-1
through Type-4, and a confidence score $\sigma\in[0,1]$.

\noindent\textbf{Goal.} Maximize detection F1 for true clone relationships
across all four clone types, with particular emphasis on Type-3 and Type-4
semantic clones that syntactic detectors miss.

We adapt the classical clone type taxonomy~\cite{roysurvey} to multi-modal
skills: Type-1 denotes identical content across all channels after whitespace
normalization; Type-2 denotes cosmetic changes such as variable renaming in
code or minor metadata edits; Type-3 denotes structural modifications including
added or removed instructions; and Type-4 denotes skills with the same
functional intent but substantially different implementation.

\subsection{Approach Overview}
\label{subsec:overview}

\begin{figure}[t]
\centering
\includegraphics[width=\columnwidth]{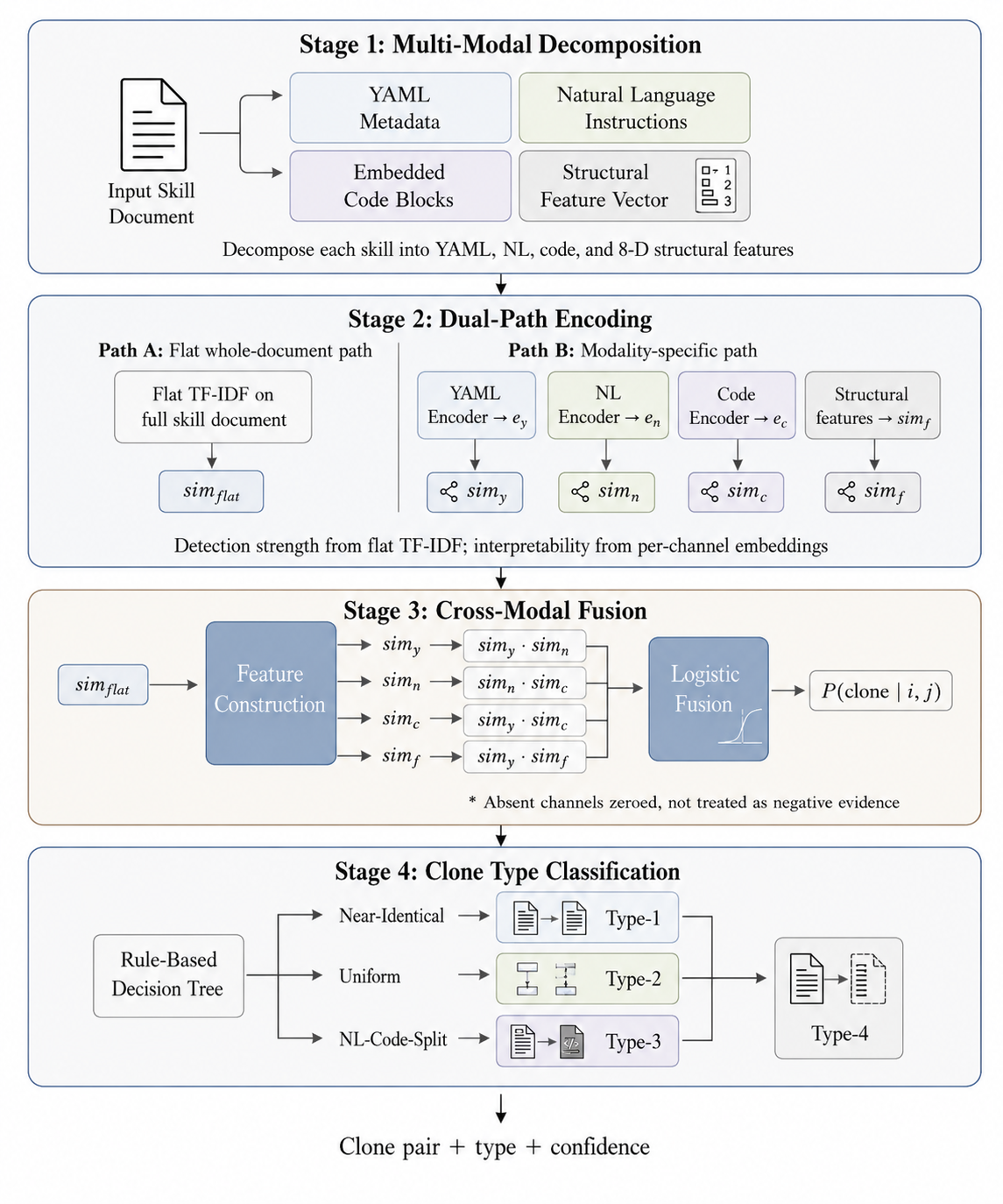}
\caption{\revision{Overview of \tool. Each skill is decomposed into four
channels, encoded via two paths (flat TF-IDF for detection strength,
per-channel TF-IDF+LSA for interpretability), fused into a calibrated clone
probability, and classified into clone types via a decision tree.}}
\Description{SkillReuse processes each skill through four stages: decomposition into YAML, natural-language, code, and structural channels; flat and channel-specific encoding; cross-modal feature fusion with logistic regression; and deterministic clone-type classification.}
\label{fig:approach}
\end{figure}

Figure~\ref{fig:approach} illustrates our four-phase approach. The design
reflects two principles. First, \emph{modality awareness}: each content channel
is encoded with techniques suited to its semantics rather than treating the
skill as flat text. Second, \emph{cross-channel interaction}: rather than
scoring channels independently and taking a weighted sum, the fusion mechanism
models pairwise interactions between channel similarities, capturing patterns
such as ``high YAML and moderate NL but absent code'' that indicate structural
adaptation (as in Figure~\ref{fig:example}).

\subsection{Multi-Modal Decomposition}
\label{subsec:decomposition}

Each skill $s_i$ is decomposed into a quadruple $s_i = (y_i, n_i, c_i,
\mathbf{f}_i)$: the YAML metadata $y_i$ extracted from the frontmatter
(fields: \texttt{name}, \texttt{description}, \texttt{category},
\texttt{tags}); the natural language instructions $n_i$ comprising the
behavioral specification after the frontmatter, with code fences removed; the
code component $c_i = \{b_1, \ldots, b_k\}$ collecting all fenced code blocks
with their language annotations; and a structural feature vector $\mathbf{f}_i
\in \mathbb{R}^8$ capturing document-level statistics.

The decomposition uses regex-based parsing to identify YAML frontmatter
(delimited by \texttt{---}), code fences (delimited by \texttt{```}), and the
remaining NL body. We filter boilerplate entries by removing skills with fewer
than five meaningful tokens after stopword removal.

The structural feature vector $\mathbf{f}_i$ captures eight document-level
properties: total word count, number of code blocks, code-to-text ratio,
average code block length, number of unique programming languages, presence of
YAML frontmatter, number of YAML fields, and description length. These features
capture the \emph{shape} of a skill document, because two skills with similar
structure (e.g., same number of code blocks, similar code fraction) are more
likely to be clones than skills with different structural profiles. Each
feature is min-max normalized across the corpus.

\subsection{Modality-Specific Encoding}
\label{subsec:encoding}

We encode each channel using latent semantic analysis
(LSA)~\cite{deerwesterlsa}: channel-specific TF-IDF
vectorization~\cite{manningir} followed by truncated SVD projection to
$d = 256$ dimensions, producing dense embeddings that capture semantic
similarity within each modality. This design follows established practices in
both NL document similarity~\cite{deerwesterlsa,manningir} and code clone
detection~\cite{sajnanisourcerercc,roysurvey}, adapting them to the
multi-modal skill setting. For the NL channel, TF-IDF+LSA captures topical
similarity in a manner analogous to sentence embedding
approaches~\cite{reimerssentence}, while avoiding external API dependencies.
For the code channel, our language-aware tokenization follows the token-based
tradition of CCFinder~\cite{kamiyaccfinder} and
SourcererCC~\cite{sajnanisourcerercc}, extended with cross-language
normalization for the multi-language code blocks found in skills.

This approach provides three advantages over neural encoders for our setting:
it requires no API calls or GPU resources, enabling full reproducibility; it
naturally handles the heterogeneous vocabulary across channels (YAML key-value
pairs, free-form text, multi-language code); \revision{and its encoding stage scales
linearly with corpus size. Neural alternatives such as
CodeBERT~\cite{fengcodebert}, GraphCodeBERT~\cite{guographcodebert}, and
UniXcoder~\cite{guounixcoder} achieve strong results on source-code clone
benchmarks, but their source-code-centric inputs do not directly represent
skills that interleave YAML, prose, and code in multiple languages.}

\noindent\textbf{YAML Channel.} We serialize the YAML metadata by
concatenating key-value pairs and project through TF-IDF+SVD to produce
$\mathbf{e}^y_i \in \mathbb{R}^d$, capturing taxonomic relatedness.

\noindent\textbf{NL Channel.} The natural language body is tokenized
with standard preprocessing (lowercasing, stopword removal) and encoded
through TF-IDF+SVD, producing $\mathbf{e}^n_i \in \mathbb{R}^d$ that captures
semantic similarity between behavioral specifications in the latent space.

\noindent\textbf{Code Channel.} Code blocks require specialized
tokenization: (1) camelCase/snake\_case splitting, (2) operator token
extraction preserving semantically meaningful operators (\texttt{|>},
\texttt{=>}), and (3) language tag prefixing
(\texttt{\_\_lang\_python\_\_}) enabling cross-language discrimination. The
resulting tokens are encoded with TF-IDF+SVD, yielding
$\mathbf{e}^c_i \in \mathbb{R}^d$.

\subsection{Cross-Modal Similarity Fusion}
\label{subsec:fusion}

The three per-channel embeddings and the structural features capture
complementary facets of skill relatedness. Following the principle that
multi-modal fusion outperforms any single
modality~\cite{liangfoundations,radfordlearning}, \revision{\tool{} uses a fusion
mechanism that combines per-channel
similarity with learned \emph{cross-channel interaction} terms, producing a
calibrated clone probability rather than an ad hoc score.}

The IDOR clone from Figure~\ref{fig:example} illustrates the central fusion
challenge. In that pair, $s_j$ contains zero code blocks, yielding $\text{sim}_c = 0.00$.
A fixed weighted sum that includes this zero dilutes the strong YAML
(${\sim}0.95$) and NL (${\sim}0.60$) signals, risking a false negative. More
subtly, the \emph{combination} of high YAML and moderate NL is stronger
evidence than either signal alone: metadata identity plus partial instruction
overlap indicates deliberate adaptation, not coincidence. The fusion mechanism
must therefore (i)~adapt to absent channels and (ii)~model interactions
between present channels.

\noindent\textbf{Intra-Channel Similarity.}
For each candidate pair $(s_i, s_j)$, we compute per-channel cosine
similarities within each modality's latent space:
\begin{equation}
\text{sim}_m(i,j) = \cos(\mathbf{e}^m_i, \mathbf{e}^m_j), \quad m \in \{y, n, c\}
\label{eq:channel}
\end{equation}
Each similarity is defined only when both skills have non-trivial content in
channel~$m$; we formalize this below.

\noindent\textbf{Similarity Profile.}
We collect the per-channel similarities into a single vector that
characterizes the pair's multi-modal relationship:
\begin{equation}
\mathbf{s}(i,j) = \bigl[\text{sim}_y(i,j),\; \text{sim}_n(i,j),\;
\text{sim}_c(i,j)\bigr] \in [0,1]^3
\label{eq:profile}
\end{equation}
The similarity profile is the primary representation for downstream clone
detection: its geometry directly encodes clone type (Type-1 pairs cluster
near $(1,1,1)$; Type-4 pairs exhibit high $\text{sim}_n$ with low
$\text{sim}_c$), making it a richer object than any scalar summary.

\noindent\textbf{Channel Presence.}
Not every skill contains every modality. We define a binary presence
indicator per channel:
\begin{equation}
P_m(i,j) = \begin{cases}
1 & \text{if channel } m \text{ is non-trivially present in both } s_i, s_j \\
0 & \text{otherwise}
\end{cases}
\label{eq:presence}
\end{equation}
When $P_m = 0$, the corresponding $\text{sim}_m$ is set to zero. Critically,
a missing channel is \emph{absence of evidence}, not evidence of dissimilarity.
The logistic model (below) learns this distinction from training pairs that
include both present and absent channels. In the IDOR example,
$P_c(i,j) = 0$ because $s_j$ has zero code blocks.

\noindent\textbf{Quadratic Feature Map.}
A simple weighted sum of per-channel similarities cannot model interactions:
it treats $\text{sim}_y = 0.95$ and $\text{sim}_n = 0.60$ as independent
additive contributions, missing the fact that their \emph{conjunction}
provides disproportionately strong evidence of cloning. We therefore lift
the similarity profile into a quadratic feature space:
\begin{equation}
\varphi(\mathbf{s}) = \bigl[\underbrace{s_y,\; s_n,\; s_c\vphantom{\big|}}_{\text{main effects}},\;\;
\underbrace{s_y {\cdot} s_n,\;\; s_n {\cdot} s_c,\;\; s_y {\cdot} s_c\vphantom{\big|}}_{\text{cross-channel interactions}}\bigr]
\in \mathbb{R}^6
\label{eq:featuremap}
\end{equation}
where $s_m$ abbreviates $\text{sim}_m(i,j)$, and any term involving a
channel with $P_m = 0$ is zeroed. The six features decompose into two groups.
The \emph{main effects} capture per-channel evidence: high $s_n$ alone
suggests topical similarity. The \emph{interaction terms} capture
cross-channel correspondence: a high $s_n {\cdot} s_c$ product indicates that
both the behavioral specification \emph{and} the implementation are similar,
providing stronger clone evidence than either channel independently. This
product-of-similarities formulation replaces the cross-space cosine used in
prior multi-modal fusion
designs~\cite{liangfoundations}: computing
$\cos(\mathbf{e}^n_i, \mathbf{e}^c_j)$ across different LSA subspaces is
geometrically ill-defined because the NL and code projections occupy
incomparable latent dimensions. The interaction term $s_n \cdot s_c$ achieves
the same goal---capturing NL-Code correspondence---through well-defined
within-space similarities.

When $P_c = 0$, the features $s_c$, $s_y {\cdot} s_c$, and
$s_n {\cdot} s_c$ all vanish, and the model reduces to three active features:
$[s_y,\; s_n,\; s_y {\cdot} s_n]$. The learned coefficients for these
features absorb the predictive burden, providing principled degradation rather
than dilution through a zero-valued channel.

\noindent\textbf{TF-IDF Base Score.}
In addition to per-channel similarities, we compute a flat TF-IDF cosine
similarity on the full \texttt{SKILL.md} content treated as a single string.
This flat score naturally captures cross-channel vocabulary overlap---a skill
about IDOR testing will contain ``idor'' in its YAML name, NL body, and code
comments. This provides a strong detection baseline that
channel-specific encoding cannot replicate. The flat score is prepended to
the feature vector as the first element.

\noindent\textbf{Structural Similarity.}
The structural feature vector $\mathbf{f}_i$
(Section~\ref{subsec:decomposition}) captures document-level shape. Its
similarity
$\text{sim}_f(i,j) = 1 - \|\mathbf{f}_i - \mathbf{f}_j\|_1 / \dim(\mathbf{f})$
is appended as the final feature, capturing whether two skills have similar
structure (code block count, code fraction, description length) regardless of
content.

\noindent\textbf{Logistic Fusion.}
The complete feature vector concatenates the flat TF-IDF score, the
quadratic channel features, and structural similarity into an 8-dimensional
input:
\begin{equation}
P(\text{clone} \mid i,j) = \sigma\!\bigl(\mathbf{w}^\top
[\,\text{sim}_{\text{flat}},\;\; \varphi(\mathbf{s}),\;\;
\text{sim}_f\,] + b\bigr)
\label{eq:fusion}
\end{equation}
where $\sigma(z) = 1/(1 + e^{-z})$ is the logistic sigmoid, and
$\mathbf{w} \in \mathbb{R}^8$, $b \in \mathbb{R}$ are learned by maximum
likelihood on annotated pairs from \bench{} with balanced class weights.
\revision{The nine parameters (eight weights and a bias) are fitted by maximum
likelihood; performance is evaluated using 5-fold stratified
cross-validation.}

This dual-path design is deliberate. The flat TF-IDF score provides a strong
detection baseline---it achieves 0.881~F1 alone
(Table~\ref{tab:rq1})---while the channel decomposition adds discriminative
evidence unavailable to flat text: interaction terms capture cross-channel
correspondence, while structural similarity captures agreement in document shape.
The logistic model learns to combine both paths, achieving 0.939~F1 by
exploiting their complementarity.

\noindent\textbf{Interpretability.}
The logistic model retains the per-channel transparency of a weighted sum
while adding interaction-level insight. For any flagged pair, the linear
combination $\mathbf{w}^\top \varphi(\mathbf{s})$ decomposes into six
additive contributions, each corresponding to a named feature. A pair whose
dominant contribution is $w_{s_n} \cdot s_n$ indicates NL-driven similarity
(possible Type-4 semantic clone); a pair dominated by
$w_{s_y \cdot s_n} \cdot s_y \cdot s_n$ indicates metadata-and-instruction
alignment without code evidence---the fractal adaptation pattern of
Figure~\ref{fig:example}. This decomposition is valuable for security and
governance applications where understanding the \emph{nature} of a clone
relationship matters as much as detecting it.



\subsection{Clone Type Classification}
\label{subsec:classification}

\revision{Once the logistic fusion identifies a clone
($P(\text{clone}) \geq \theta$), the similarity profile
$\mathbf{s} = [s_y, s_n, s_c]$ is used to assign its clone \emph{type}.}
Rather than partitioning the profile space with
ad hoc score ranges, we define three Boolean predicates that each encode a
structural property from the classical clone taxonomy
(Table~\ref{tab:clonetypes}), then compose them into a deterministic decision
tree.

\revision{Clone detection and type assignment are separate stages. Detection estimates whether a pair is a clone and therefore uses the complete 8-dimensional vector from Section~\ref{subsec:fusion}: the flat score, three per-channel similarities, three pairwise interactions, and structural similarity. After detection, the clone type is not predicted by a second model; it is assigned deterministically by applying the adapted taxonomy to the three per-channel similarities $(s_y,s_n,s_c)$. These scores describe how similarity is distributed across YAML, NL, and code, which is the distinction the taxonomy encodes. The remaining five features support clone detection but provide no additional information required by the taxonomy: the flat and structural scores do not identify which channels changed, while the interaction terms are derived from the three channel scores and introduce no independent channel-level condition. Consequently, a feature that contributes strongly to detection need not appear in the subsequent type-assignment rule. We therefore use the assigned types as heuristic taxonomic descriptions rather than learned predictions. RQ1 evaluates binary detection, using the benchmark's ground-truth types only to stratify recall, and RQ3 constructs the ecosystem graph solely from binary clone decisions; neither analysis depends on the assigned type labels.}

\noindent\textbf{Predicates.}
Let $\mathcal{P}$ denote the set of channels present in both skills and
$\mathbf{s}_\mathcal{P}$ the corresponding sub-vector.
\begin{itemize}[nosep,leftmargin=*]
\item \textsc{Near-Identical}$(\mathbf{s})$:
  $\;\min(\mathbf{s}_\mathcal{P}) \geq \tau_1$.
  All present channels are near-verbatim---the defining property of Type-1
  (``exact except layout/comments'').
\item \textsc{Uniform}$(\mathbf{s})$:
  $\;\max(\mathbf{s}_\mathcal{P}) - \min(\mathbf{s}_\mathcal{P}) \leq \delta$.
  Channel similarities agree within tolerance~$\delta$---structure is
  preserved, consistent with Type-2 (``cosmetic relabeling'').
\item \textsc{NL-Code-Split}$(\mathbf{s})$:
  $\;s_n \geq \tau_2 \;\wedge\; (s_c < \tau_2 \;\vee\; P_c {=} 0)$.
  The intent channel (NL) shows similarity but the implementation channel
  (code) diverges or is absent---the defining signature of Type-4
  (``same semantics, different syntax'').
\end{itemize}

\noindent\textbf{Decision Tree.}
The three predicates compose into a four-leaf tree with no overlapping
regions:
\begin{enumerate}[nosep,leftmargin=*]
\item If \textsc{Near-Identical}$(\mathbf{s})$: \textbf{Type-1}.
\item Else if \textsc{Uniform}$(\mathbf{s})$: \textbf{Type-2}.
\item Else if \textsc{NL-Code-Split}$(\mathbf{s})$: \textbf{Type-4}.
\item Else: \textbf{Type-3} (asymmetric profile with partial channel edits).
\end{enumerate}
The ordering is deliberate: Type-1 is a strict subset of Type-2 (all
near-identical profiles are also uniform), so it must be tested first. Type-4
is tested before the Type-3 fallback because the NL-Code split is a specific
pattern of asymmetry that warrants its own category.

\noindent\textbf{Parameters.}
The tree requires only three values: $\tau_1$ (identity threshold),
$\delta$ (uniformity tolerance), and $\tau_2$ (semantic threshold).
\revision{$\tau_1 = 0.90$ follows Roy and Cordy's standard operationalization
of near-verbatim Type-1 clones~\cite{roysurvey}, which we adopt rather than
re-deriving from our data. $\delta$ and $\tau_2$ have no equivalent external
anchor, because they encode properties specific to the multi-channel setting
(cross-channel agreement and the NL-code split), so we fix them a priori as
interpretable defaults: $\delta = 0.20$ is the tolerance within which channel
similarities count as uniform, and $\tau_2 = 0.50$ places the Type-4 boundary
at the natural high/low midpoint. These are fixed heuristic boundaries layered
on top of the logistic detector's clone decision, not tuned on data and not
independently learned evidence. They affect only the interpretive type label:
the per-type recall in Section~\ref{sec:evaluation} buckets detected pairs by
ground-truth type and does not depend on them, and
Section~\ref{sec:discussion} treats the labeling itself as a threat to
validity.}

\noindent\textbf{Example.}
The IDOR pair from Figure~\ref{fig:example} has
$\mathbf{s} = [0.95,\; 0.60,\; 0.00]$ with $P_c = 0$.
\textsc{Near-Identical} fails ($\min = 0.60 < 0.90$).
\textsc{Uniform} fails ($0.95 - 0.60 = 0.35 > 0.20$).
\textsc{NL-Code-Split}: $s_n = 0.60 \geq 0.50$ and $P_c = 0$ $\Rightarrow$
true, so the pair is classified as \textbf{Type-4}. The decision trace is
fully auditable: each predicate maps to a named property from the clone
taxonomy.

\section{Evaluation}
\label{sec:evaluation}

We evaluate \tool to answer four research questions:
\begin{description}[leftmargin=0pt, style=unboxed]
 \item[\RQ{1} (Detection Effectiveness):] How effective is \tool at detecting
 clone relationships overall and across clone types, compared to
 single-channel and token-based baselines?
 \item[\RQ{2} (Ablation):] How does each modality channel and the cross-modal
 alignment contribute to detection performance?
 \item[\RQ{3} (Ecosystem Analysis):] What clone patterns does \tool reveal at
 ecosystem scale?
 \item[\RQ{4} (\revision{Security-Relevant Blast Radius}):] \revision{How far does the clone graph extend from skills containing security-relevant patterns?}
\end{description}

\subsection{Registry Corpus}
\label{sec:corpus}

We crawled the skillsmp.com~\cite{skillsmp} registry, retrieving \texttt{SKILL.md} content for
155,547 of 196,134 listed skills (79.3\% success rate; failures due to deleted
or private repositories). The corpus totals 1.19\,GB (avg.\ 7.6\,KB per
skill); 78\% of skills contain code blocks and 43\% are code-heavy
($>$50\% code), confirming the need for multi-modal analysis.
\revision{We refer to these 155,547 retrieved files as the content-bearing corpus. After applying the five-meaningful-token content filter from Section~\ref{subsec:decomposition}, 137,470 skills remain in the analysis corpus; RQ3 uses all of them, with no sampling.}

\subsection{Benchmark Construction}
\label{sec:benchmark}

No benchmark exists for clone detection on multi-modal agent skills.
We construct \bench, a balanced benchmark of 300 ground-truth pairs (150
positive, 150 negative) with stratified difficulty across both classes.

\noindent\textbf{Design Principles.}
We prioritize quality over quantity: each pair is included for a specific
evaluation purpose, and both classes are stratified by difficulty so that
aggregate metrics are not dominated by trivial cases. Positives span four
clone types at increasing difficulty; negatives range from obvious non-matches
to deliberately confusing same-category pairs. \revision{We apply objective,
auditable quality controls based on repository provenance, byte-level hashes,
controlled mutations, and constrained negative sampling.}

\revision{\noindent\textbf{Label Quality Controls.} We use three auditable criteria.
\textbf{(1) Fork-mined positives:} each pair must have an explicit GitHub fork relationship;
we additionally record SHA-256 hashes to distinguish byte-identical copies from
modified forks. \textbf{(2) Synthetic positives:} each pair is generated from a
real seed skill by exactly one of the seven deterministic mutation operators
M1--M7, so its transformation is known by construction. \textbf{(3) Negatives:}
each pair must have no SHA-256 overlap with any positive seed; the pairs are
then stratified into cross-category
(easy) and same-category (hard) pools. The hard pool therefore shares domain
vocabulary and cannot be separated by topic words alone. Each benchmark entry
is traceable to a fork relationship, a named mutation operator, or a documented
hash/category constraint. Released scripts support independent audit and extension.}

\noindent\textbf{Positive Pairs (150).}
We draw equally from two complementary sources, each contributing 75 pairs:

\emph{Fork-mined pairs} (75) exploit GitHub fork relationships as verifiable
provenance. When a developer forks a skill repository and modifies its
contents, the fork tree confirms derivation. We match skills across forks by
file path and name similarity, then stratify by modification level into
Type-1 (verbatim), Type-2 (renamed), Type-3 (structurally modified), and
Type-4 (semantically equivalent but rewritten).

\emph{Synthetic mutations} (75) ensure coverage of controlled difficulty
levels via seven operators applied to randomly sampled seed skills:
M1~(identifier renaming, Type-2),
M2~(NL paraphrasing with synonym replacement and 30\% content removal, Type-3),
M3~(code stripping that retains only NL section headers, Type-4),
M4~(content extension, Type-3),
M5~(content subsetting, Type-3),
M6~(cross-skill combination, Type-3), and
M7~(full semantic rewrite via LLM paraphrase, Type-4).
M3 and M7 produce the hardest pairs: both preserve functional intent while
eliminating surface-level token overlap.

\noindent\textbf{Negative Pairs (150).}
To avoid easy-negative inflation, we stratify negatives by difficulty:

\emph{Easy negatives} (50) pair skills from different categories (e.g.,
DevOps vs.\ documentation), providing clear non-clones where low similarity
is expected.

\emph{Hard negatives} (100) pair skills from the \emph{same} category whose
multi-modal similarity falls in $[0.15, 0.45]$. These pairs share topical
vocabulary (e.g., two Kubernetes skills both referencing \texttt{kubectl} and
\texttt{namespace}) without being clones. Hard negatives constitute two-thirds
of the negative set, deliberately stressing the detector's ability to
distinguish vocabulary overlap from genuine cloning.

\begin{table}[t]
\centering
\caption{\bench composition: 300 pairs, balanced 150/150 and stratified by
difficulty within each class.}
\label{tab:bench}
\begin{minipage}{\columnwidth}
\centering
\begin{tabularx}{\columnwidth}{@{}Xrrr@{}}
\toprule
\textbf{Source} & \textbf{Pos.} & \textbf{Neg.} & \textbf{Total} \\
\midrule
Fork-mined & 75 & --- & 75 \\
Synthetic mutation & 75 & --- & 75 \\
Easy negatives (cross-category) & --- & 50 & 50 \\
Hard negatives (same-category) & --- & 100 & 100 \\
\midrule
\textbf{Total} & \textbf{150} & \textbf{150} & \textbf{300} \\
\bottomrule
\end{tabularx}

\end{minipage}
\end{table}

Table~\ref{tab:bench} summarizes the composition. Within positives,
\emph{easy} pairs (Type-1/2) are near-verbatim copies detectable by any
approach, while \emph{hard} pairs (Type-3/4, 55\% of positives) involve
structural modifications or semantic rewrites that challenge single-channel
detectors. Within negatives, \emph{easy} pairs cross category boundaries
(e.g., DevOps vs.\ documentation), while \emph{hard} pairs share the
\emph{same} category and topical vocabulary (e.g., two Kubernetes skills)
without being clones. Hard negatives constitute two-thirds of the negative
set, deliberately stressing the detector's false-positive rate. We release
\bench{} as part of our replication package.

\subsection{Experimental Setup}

\noindent\textbf{Baselines.} \revision{We compare against two methods
independent of \tool{} (TF-IDF, MinHash) and two single-channel restrictions
of \tool{} itself (NL-Only, Code-Only), kept distinct from the RQ2
 feature-ablations of \tool{}'s own model in Table~\ref{tab:ablation}. All
 four use the same corpus and address the same binary clone-detection task:}
\begin{itemize}[leftmargin=*]
 \item \textbf{TF-IDF (flat text):} Standard TF-IDF cosine similarity on the
 full \texttt{SKILL.md} content as a single string, ignoring modality
 structure.
 \item \textbf{MinHash:} Locality-sensitive hashing on character $n$-grams
 ($n\!=\!5$) of the full skill text.
 \item \textbf{NL-Only:} \tool's NL channel embedding applied in isolation,
 representing the strongest single-channel baseline.
 \item \textbf{Code-Only:} \tool's code channel embedding applied in
 isolation.
\end{itemize}

\noindent\textbf{Baseline Scope.} Our goal is to compare \tool against
baselines that preserve the task definition, namely clone detection over full
multi-modal skill documents rather than over extracted code fragments alone. We
therefore do not present this evaluation as an exhaustive state-of-the-art
bakeoff against every recent code-clone model. Instead, we compare against
strong text and token baselines plus \revision{single-channel restrictions of \tool{}}
	that operate on the same underlying artifacts. \revision{We do not include neural code clone detectors
	such as DeepSim~\cite{zhaodeepsim} or
	MAGNET~\cite{magnetclone} because their source-code-centric representations
	do not directly accept heterogeneous \texttt{SKILL.md} artifacts. Applying these tools
would require discarding the NL and YAML channels (78\% of skill content by
volume), fundamentally changing the detection task.} The NL-Only and Code-Only
\revision{rows} in Table~\ref{tab:rq1} quantify the cost of this single-channel
restriction: both achieve lower F1 than multi-modal fusion.

\noindent\textbf{Metrics.} We report precision, recall, and F1 score for
binary clone detection. For each approach, we find the optimal threshold via
grid search on \bench to ensure a fair comparison at each method's best
operating point. For clone type analysis, we report per-type recall.

\noindent\textbf{Environment.} All experiments run on a single machine with
32~GB RAM and an Intel Xeon CPU. We implement \tool and baselines in Python, using scikit-learn for TF-IDF and logistic regression. 

\subsection{RQ1: Detection Effectiveness}

\begin{table}[t]
\centering
\caption{Clone detection on \bench (300 pairs).
Per-type columns show recall; per-type precision is undefined because false positives lack
ground-truth types.}
\label{tab:rq1}
\setlength{\tabcolsep}{2pt}
\begin{tabular}{@{}lrrr@{\;\;}rrrr@{}}
\toprule
 & \multicolumn{3}{c}{\textbf{Overall}} & \multicolumn{4}{c}{\textbf{Per-Type Recall}} \\
\cmidrule(lr){2-4} \cmidrule(l){5-8}
\textbf{Approach} & \textbf{P} & \textbf{R} & \textbf{F1}
  & \textbf{T1} & \textbf{T2} & \textbf{T3} & \textbf{T4} \\
 & & & & (18) & (25) & (50) & (57) \\
\midrule
TF-IDF (flat)      & .897 & .867 & .881  & 1.00 & 1.00 & .98 & .67 \\
MinHash             & .962 & .673 & .792  & 1.00 & 1.00 & .94 & .19 \\
NL-Only             & .814 & .847 & .830  & 1.00 & 1.00 & 1.00 & .60 \\
Code-Only           & .802 & .673 & .732  & .83  & .96  & .78 & .40 \\
\textbf{\tool}      & \textbf{.952} & \textbf{.927} & \textbf{.939}
  & \textbf{1.00} & \textbf{1.00} & \textbf{1.00} & \textbf{.81} \\
\bottomrule
\end{tabular}
\end{table}

Table~\ref{tab:rq1} presents the results.
\tool achieves the highest F1 (0.939) with both the highest precision (0.952)
and recall (0.927). It also achieves perfect recall on Type-1--3 and the
highest Type-4 recall (81\%), outperforming TF-IDF (67\%) and MinHash (19\%).
The interaction terms in the logistic fusion capture cross-channel
correspondence that flat bag-of-words cannot model: a pair with high YAML
\emph{and} moderate NL receives a super-additive boost that TF-IDF misses.
5-fold cross-validation estimates \tool's generalization F1 at 0.920
($\pm$0.026).

MinHash achieves high precision (0.962) but sacrifices recall (0.673),
missing one-third of clones. The single-channel baselines (NL-Only 0.830,
Code-Only 0.732) confirm that combining channels via logistic fusion
substantially outperforms any single modality.

\revision{\noindent\textbf{With Versus Without \tool{}.} On the 150 positive benchmark pairs, \tool{} detects 139 and misses 11, compared with 130 detected and 20 missed by the strongest F1 baseline, flat TF-IDF. \tool{} also produces 7 false positives, compared with 15 for TF-IDF. The gap is larger for scalable token matching: MinHash detects only 101 positives and misses 49. The single-channel restrictions likewise recover fewer positives (127 for NL-Only and 101 for Code-Only), showing that the improvement comes from combining evidence across the skill's heterogeneous channels rather than applying a conventional or single-channel matcher.}

\revision{\noindent\textbf{Per-Type and Error Analysis.} The per-type columns
reveal where multi-modal fusion provides the greatest advantage. \tool{}
achieves perfect recall on Types 1--3, while the clearest differentiation
emerges for Type-4 semantic clones.} \tool achieves 81\%
Type-4 recall (46/57), which is 4.2$\times$ higher than MinHash and
1.2$\times$ higher than TF-IDF. MinHash fails on Type-4 because these clones
share minimal surface tokens; TF-IDF captures some through vocabulary overlap
but cannot model cross-channel interactions. \revision{The 95\% Wilson intervals
for \tool{} are Type-1 18/18 [0.824,1.000], Type-2 25/25 [0.867,1.000],
Type-3 50/50 [0.929,1.000], and Type-4 46/57 [0.687,0.889]. Even at its
lower bound, \tool{}'s Type-4 interval is 37.4pp above MinHash's
[0.111,0.313].}

\revision{\noindent\textbf{On the Type Label.} The per-type recall
above measures whether the detector flags a pair as a clone; the type label
itself is a separate, interpretive layer. As explained in
Section~\ref{subsec:classification}, we assign the type directly from the
per-channel similarity profile, which is what the classical taxonomy is
defined over, and present it as an interpretable description of the dominant
similarity pattern of a detected clone, not as a calibrated per-pair
classifier. All of the paper's quantitative claims, RQ1 detection and the
RQ3 ecosystem analysis alike, rest on the binary detector, not on the
assigned type.}

\tool produces 11 false negatives and 7 false positives on \bench. All 11
false negatives are Type-4 pairs where both NL and code differ substantially
despite shared functional intent, making them the hardest cases at the
boundary of what structural similarity can detect. Among false positives, the
dominant category is same-category hard negatives where shared domain
vocabulary (e.g., two Kubernetes skills with \texttt{kubectl} tokens) inflates
similarity across channels.

\begin{tcolorbox}[colback=gray!10,colframe=gray!50,boxrule=0.5pt]
\textbf{Answering RQ1:} \tool achieves the highest F1 (0.939), precision
(0.952), and recall (0.927), outperforming flat TF-IDF by 5.8pp~F1. It also
achieves the best per-type \revision{detection} recall, including 81\% Type-4
recall (vs.\ 67\% for TF-IDF and 19\% for MinHash); all missed clones are
hard Type-4 semantic pairs.
\end{tcolorbox}

\revision{\noindent\textbf{Benchmark Scale-Robustness.} To confirm
these results are not an artifact of benchmark size, we expand \bench{} by
50\% to 450 pairs, following the same construction protocol
(Section~\ref{sec:benchmark}) and evaluating every method under the same setup.
\tool{} remains the strongest method: F1 0.886 (precision 0.888, recall 0.884),
a decrease of 5.3 percentage points from 0.939 on the 300-pair set, while still
outperforming the strongest baseline at 0.850. Types 1--3 retain perfect recall with tighter 95\%
Wilson intervals (Type-1 [0.896, 1.000], Type-2 [0.890, 1.000], Type-3 [0.939,
1.000]); the small overall dip concentrates in Type-4, whose recall moves from
0.807 to 0.745 (Wilson [0.653, 0.820]) as the enlarged same-category pool
introduces harder semantic-clone pairs. The modest F1 decrease and retained
	lead over all baselines indicate that performance is largely preserved as the
	benchmark grows and becomes more difficult, providing evidence of robustness
	across benchmark scale and difficulty. False positives follow the
same-category vocabulary-overlap mode noted above.}

\subsection{RQ2: Ablation Study}

\begin{table}[t]
\centering
\caption{\revision{Ablation study. Each variant removes the indicated feature
group(s) and retrains the logistic model on the remaining features.}}
\label{tab:ablation}
\begin{tabular}{@{}lrrr@{}}
\toprule
\textbf{Configuration} & \textbf{F1} & \textbf{$\Delta$F1} \\
\midrule
\tool (full model) & 0.939 & --- \\
\quad \revision{No channel features (TF-IDF only)} & 0.878 & $-$6.1pp \\
\quad \revision{No structural features} & 0.890 & $-$4.9pp \\
\quad \revision{No TF-IDF base} & 0.907 & $-$3.2pp \\
\quad \revision{No YAML channel} & 0.928 & $-$1.1pp \\
\quad \revision{No code channel} & 0.929 & $-$1.0pp \\
\bottomrule
\end{tabular}
\end{table}

Table~\ref{tab:ablation} validates that every component contributes.
The largest single-component impact comes from channel decomposition:
removing all channel features (``TF-IDF only'' row) costs $-$6.1pp,
confirming that per-channel similarity and interaction terms provide
substantial evidence beyond flat document similarity. Structural features
contribute $-$4.9pp, reflecting the discriminative value of document-level
shape (code block count, code fraction) for distinguishing clones from
same-category non-clones. The TF-IDF base contributes $-$3.2pp, confirming
its role as a \revision{strong detection signal}. The YAML and code channels each
contribute approximately $-$1pp individually.

\begin{tcolorbox}[colback=gray!10,colframe=gray!50,boxrule=0.5pt]
\textbf{Answering RQ2:} Channel decomposition provides the largest
contribution ($-$6.1pp when removed), followed by structural features
($-$4.9pp) and TF-IDF base ($-$3.2pp). All components contribute positively;
the full fusion outperforms TF-IDF alone by 6.1pp.
\end{tcolorbox}

\subsection{RQ3: Ecosystem-Scale Analysis}
\label{sec:ecosystem}

\revision{We run the full \tool{} pipeline on all 155,547 skills in the content-bearing corpus. Its document-level content filter retains 137,470 eligible skills for pairwise clone analysis, all of which are included with no sampling. Every number in this subsection is regenerated by a single committed script over this analysis corpus, so the analysis is reproducible end-to-end rather than dependent on a particular sample.}

\revision{\noindent\textbf{Prefilter.} After the document-level content filter,
the analysis corpus contains $N=137{,}470$ skills and approximately 9.4 billion
unordered pairs. To keep this $O(n^2)$ candidate set tractable, we use dense
NL-channel LSA embeddings and batched cosine-similarity computation to retain
approximately 5.0 billion pairs (53\%) at
$\theta_{\text{cand}}=0.10$ for full multi-modal scoring. Applied
retroactively to \bench{}, the prefilter retains 148/150 positives (98.7\%
candidate recall). We therefore use it only for ecosystem-scale candidate
generation, not as a second clone decision.}

\begin{table}[t]
\centering
\caption{\revision{Ecosystem-scale clone and deduplication statistics over
the full 137,470-skill analysis corpus, at our conservative operating
point ($\theta{=}0.80$).}}
\label{tab:ecosystem}
\begin{tabular}{@{}lr@{}}
\toprule
\textbf{Metric} & \textbf{Value} \\
\midrule
Skills analyzed & \revision{137,470} \\
Exact duplicates (SHA-256) & \revision{20,016 (14.6\%)} \\
Skills in $\geq$1 clone pair & \revision{91,792 (66.8\%)} \\
Clone pairs detected & \revision{1,059,923} \\
Cross-author clone pairs & \revision{1,010,480 (95.3\%)} \\
\midrule
\multicolumn{2}{@{}l}{\textit{Deduplication:}} \\
\quad After exact dedup & \revision{117,454} \\
\quad \revision{Estimated skill concepts (components)} & \revision{56,043} \\
\quad \textbf{Ecosystem inflation ratio} & \revision{\textbf{2.45$\times$}} \\
\midrule
\multicolumn{2}{@{}l}{\textit{Cluster structure:}} \\
\quad \revision{Singleton components} & \revision{45,678} \\
\quad Multi-skill clusters & \revision{10,365} \\
\quad Largest cluster & \revision{61,651 skills} \\
\bottomrule
\end{tabular}
\end{table}

Table~\ref{tab:ecosystem} summarizes the ecosystem-level findings.

\revision{\noindent\textbf{With Versus Without \tool{}.} Without skill-aware clone detection, exact SHA-256 deduplication removes 20,016 redundant skill entries and leaves 117,454 distinct files, corresponding to only 1.17$\times$ inflation. \tool{} instead groups the same 137,470 analyzed skills into 56,043 concepts, exposing 81,427 redundant skill entries and 2.45$\times$ inflation. Under these matched definitions, exact hashing captures only 24.6\% of the redundancy exposed by \tool{} and leaves 61,411 additional redundant skill entries unresolved. Thus, conventional exact deduplication substantially understates ecosystem reuse.}

\revision{\noindent\textbf{Threshold Sensitivity.} The effect of the detection threshold $\theta$ differs substantially across metrics; we therefore distinguish conclusions that remain stable across the threshold sweep from those that require qualification.
The exact SHA-256 duplicate rate is independent of $\theta$ (14.6\%), while the cross-author share of clone links varies from 96.5\% at $\theta{=}0.60$ to 95.3\% at $\theta{=}0.80$.
Clone involvement is moderately sensitive, decreasing from 89.0\% to 66.8\% over the same range ($1.3\times$).
We report the value at $\theta{=}0.80$ as a conservative lower bound because this is the strictest threshold evaluated and maintains approximately 98\% precision on \bench{} (0.983 on the original 300-pair benchmark and 0.988 on its 450-pair expansion).
The inflation ratio is the most threshold-sensitive metric because small
changes in the inferred edge set can merge connected components through
transitive chains. We therefore report the $2.45\times$ value at
$\theta{=}0.80$ only as a conservative derived estimate.
The complete threshold sweep is available in the replication package.}

\revision{\noindent\textbf{Complexity-Stratified Clone Rates.} To test whether
trivial stubs inflate the 66.8\% clone-involvement rate, we recompute it after
excluding every skill below 500\,B. As Table~\ref{tab:complexity} shows, stubs
comprise 2,038 of 137,470 skills (1.5\% of the analysis corpus) and only 822 of
91,792 clone-involved skills (0.9\%). Removing them raises the aggregate rate
from 66.8\% to 67.2\%; stubs therefore slightly depress rather than inflate the
headline result. The higher rates in longer bins are descriptive and are not
interpreted causally.}

\begin{table}[t]
\centering
\caption{\revision{Clone-involvement rate by content-length bin over the full 137,470-skill analysis corpus. Parentheses in the bin rows give shares of the corresponding column; ``In clones'' denotes skills in a multi-skill cluster at the conservative operating point $\theta{=}0.80$.}}
\label{tab:complexity}
\setlength{\tabcolsep}{3pt}
\begin{tabular}{@{}lrrr@{}}
\toprule
\textbf{Length bin} & \textbf{Skills} & \textbf{In clones} & \textbf{Clone Rate} \\
\midrule
\revision{$<500\,\mathrm{B}$ (stub)} & \revision{2,038 (1.5\%)} & \revision{822 (0.9\%)} & \revision{40.3\%} \\
\revision{$500\,\mathrm{B}$--$2\,\mathrm{KB}$} & \revision{22,199 (16.1\%)} & \revision{11,620 (12.7\%)} & \revision{52.3\%} \\
\revision{$2\,\mathrm{KB}$--$5\,\mathrm{KB}$} & \revision{40,846 (29.7\%)} & \revision{25,821 (28.1\%)} & \revision{63.2\%} \\
\revision{$5\,\mathrm{KB}$--$30\,\mathrm{KB}$} & \revision{69,884 (50.8\%)} & \revision{51,413 (56.0\%)} & \revision{73.6\%} \\
\revision{$>30\,\mathrm{KB}$ (large)} & \revision{2,503 (1.8\%)} & \revision{2,116 (2.3\%)} & \revision{84.5\%} \\
\midrule
\textbf{Total} & \revision{\textbf{137,470}} & \revision{\textbf{91,792}} & \revision{\textbf{66.8\%}} \\
\revision{Excluding stubs} & \revision{135,432} & \revision{90,970} & \revision{\textbf{67.2\%}} \\
\bottomrule
\end{tabular}
\end{table}

We highlight three observations that challenge common assumptions about agent skill ecosystems.

\subsubsection{Finding 1: Most Duplicates Are Modified Variants}

\revision{Only 14.6\% of skills are exact (SHA-256) duplicates, yet 66.8\%
participate in clone relationships even at our conservative threshold. Within
name-based clone families, 97\% of skill pairs are modified variants and only
3\% are byte-identical copies, showing that modified variants dominate over
verbatim copies.
This contrasts sharply with traditional package ecosystems, where reuse is
mediated by dependency declarations and verbatim copies are
rare~\cite{zimmermannsmall}. This form of reuse is invisible to the registry
infrastructure.}

\subsubsection{Finding 2: Author Concentration and Aggregator Mirroring}

\revision{Across the full analysis corpus of 14,873 authors, production is concentrated
but not extreme: the top 10 authors produce 24.7\% of all skills, at a Gini
coefficient of 0.746. This is more concentrated than npm, where the top 1\%
of maintainers account for $\sim$10\% of packages~\cite{zimmermannsmall}, but
far milder than the 20,000-skill crawl-order sample suggested, where a handful of
prolific accounts dominated the first records. The full-corpus estimate is
therefore more representative.}

\revision{Aggregator accounts are an important source of redundancy because they collect and republish skills from other authors, often creating near-identical copies under different names.
Their contribution is substantial but does not explain the overall pattern.
Excluding the single highest-volume author reduces the inflation ratio from 2.45$\times$ to 2.27$\times$ and clone involvement from 66.8\% to 61.5\%; excluding the ten highest-volume accounts still leaves 2.10$\times$ inflation and 56.1\% clone involvement.
These results indicate that aggregator mirroring accounts for part of the observed redundancy, while copy-and-modify reuse remains widespread across the rest of the ecosystem.}

\subsubsection{Finding 3: Copying Leaves Many Skills Superseded}

\revision{Copy-and-modify reuse produces not only variation but also
differences in completeness. To quantify them, we rank skills within each
name-family ($\geq$3 members, 54,270 skills across 8,405 families) by a
quality composite of content length, code-block count, and natural-language
coverage, and mark a skill \emph{superseded} when another family member
weakly dominates it on all three axes and strictly exceeds it on at least
one. By this measure, \textbf{67\% of skills in clone families (36,526 of
54,270) are superseded}: a strictly more complete variant of the same skill
already exists, often under a different author. Clone detection therefore
enables \emph{quality-aware deduplication}: rather than deleting duplicates,
a registry can surface the best variant in each cluster and link the
alternatives.}


\begin{tcolorbox}[colback=gray!10,colframe=gray!50,boxrule=0.5pt]
\textbf{Answering RQ3:} \revision{Across the full 137,470-skill corpus, 66.8\% of skills participate in clone relationships, and the clone graph reduces the corpus to 56,043 concepts, corresponding to a conservative 2.45$\times$ inflation estimate. Because 97\% of within-family clone pairs are modified rather than verbatim copies, exact hashing misses the dominant form of redundancy; moreover, 67\% of skills in name-based clone families have a more complete alternative, showing the value of clone detection for quality-aware deduplication.}
\end{tcolorbox}

\subsection{\revision{RQ4: Security-Relevant Blast Radius}}
\label{sec:security}

\revision{As a security application of the clone graph, we scan the
entire 137,470-skill corpus
for security-relevant skills using regex patterns for six
categories of code associated with standard weakness
classes~\cite{cwe}: SQL injection payloads (CWE-89), XSS test
vectors (CWE-79), reverse shell commands (CWE-78),
command injection via unsafe subprocess calls (CWE-78),
hardcoded secrets (CWE-798), and unsafe \texttt{eval} on user
input (CWE-95). The regex signatures follow established
vulnerability references: the OWASP cheat sheets~\cite{owaspxss,owaspsqli},
the PentestMonkey reverse-shell reference~\cite{pentestmonkey}, the Bandit
Python security linter~\cite{banditlinter}, and secret-scanning
heuristics~\cite{detectsecrets}; the seed derivation is a committed,
deterministic script that records the matched pattern for every seed.
We then use \tool{} to identify each seed's clone-graph neighborhood, which we
call its blast radius, at the conservative detection threshold $\tau{=}0.80$
(98\% precision on \bench) so that the neighborhoods comprise high-confidence
clones. This relation is non-directional and does not indicate which skill
came first.}

\begin{table}[t]
\centering
\caption{\revision{Clone-graph blast radius from 938 security-relevant skills
across the entire 137,470-skill corpus (RQ4).
``Retained'' = clones that also contain the seed's security-relevant pattern. Per-category rows may overlap.}}
\label{tab:security}
\resizebox{\columnwidth}{!}{%
\begin{tabular}{@{}lrrrr@{}}
\toprule
\textbf{Pattern} & \textbf{Seeds} & \textbf{Clones} & \textbf{Cross-Auth.} & \textbf{\revision{Ret.}} \\
\midrule
XSS payloads & \revision{613} & \revision{10,224} & \revision{9,852} & \revision{1,535} \\
SQL injection & \revision{232} & \revision{5,723} & \revision{5,491} & \revision{1,281} \\
Hardcoded secrets & \revision{72} & \revision{687} & \revision{612} & \revision{85} \\
Reverse shell & \revision{37} & \revision{338} & \revision{306} & \revision{163} \\
Cmd.\ injection & \revision{24} & \revision{210} & \revision{205} & \revision{21} \\
Unsafe eval & \revision{64} & \revision{1,379} & \revision{1,289} & \revision{106} \\
\midrule
\textbf{All \revision{(unique)}} & \textbf{\revision{938}} & \textbf{\revision{16,587}} & \textbf{\revision{15,886}} & \textbf{\revision{2,676}} \\
\bottomrule
\end{tabular}%
}
\end{table}

\revision{Table~\ref{tab:security} shows the results. We identify 938 seed skills containing security-relevant code, of which 760 (81\%) have at least one clone in the corpus. Their blast radius comprises 16,587 clone relationships spanning 6,376 distinct skills maintained by 2,301 authors. Of these clone links, 15,886 (96\%) connect skills held by different authors.}

\noindent\textbf{\revision{Pattern Retention}.}
\revision{Of the 16,587 clone links, 2,676 (16\%) also contain the seed's security-relevant code pattern. Here, ``retention'' is a content-level observation: it does not mean that the clone is an exact copy or that the pattern propagated from the seed. The remaining 84\% are structurally or semantically related but do not contain the pattern. For example, \texttt{davila7}'s ``Linux Privilege Escalation'' skill (reverse shell commands) has 22 clones, 21 held by other authors and 8 also containing reverse-shell code. Its ``Pentest Commands'' skill (SQL injection payloads) has 16 clones, all held by other authors and 9 also containing SQL-injection payloads.}

\noindent\textbf{Implications.}
\revision{A registry maintainer who flags and removes one
pentesting skill would miss a family of 22 clones, 8 of which also contain
reverse-shell code, under different names and authors. Without clone-graph
analysis, the potential blast radius remains hidden.}

\noindent\textbf{\revision{Confirmed Malware.}}
\revision{By decoding base64-obfuscated install commands, we identify 22 distinct skills that execute the ClawHavoc campaign's Atomic Stealer payload, verified against its published command-and-control address~\cite{clawhavoc}. These skills span three publisher accounts, and one trojanized skill recurs with an identical payload under two of them. The clone graph links the malicious skills to high-scoring relatives (0.94--0.96) that share the legitimate skill body but usually omit the injected payload; only three clone links retain it. Thus, a high clone score with a malicious skill reveals the shared underlying skill, not necessarily a copy of the malware. The graph provides a focused candidate set for security auditing.}

\begin{tcolorbox}[colback=gray!10,colframe=gray!50,boxrule=0.5pt]
\textbf{Answering RQ4:} \revision{The clone graph reveals related candidates that per-skill scanning misses. In the confirmed ClawHavoc case, high-scoring relatives shared the legitimate skill body, but only three clone links retained the payload; clone analysis therefore guides follow-up, while payload inspection establishes malware status.}
\end{tcolorbox}

\section{Discussion}
\label{sec:discussion}

\subsection{Clone Detection vs.\ Reuse Inference}

\revision{Clone detection establishes structural similarity, not derivation. Confirming that $s_j$ was derived from $s_i$ requires provenance such as fork relationships, commit timestamps, or explicit attribution; our fork-mined benchmark provides such evidence. RQ3 therefore reports \emph{clone prevalence}, not reuse prevalence: the 1.06 million detected pairs may include both intentional copying and convergent development.}

\revision{As shown in Section~\ref{sec:security}, a high clone score can reflect a shared legitimate skill body rather than a shared malicious payload. In practice, clone detection identifies structurally related artifacts, while repository history establishes provenance and payload-level inspection establishes security status. Clone edges should therefore be treated as prioritization links rather than as evidence of dependency or propagation.}

\subsection{Implications}

\revision{Our findings have two practical implications. For registry operators, the 66.8\% clone-involvement and 67\% superseded-skill rates motivate deduplication and provenance tracking as first-class features. Deduplication need not remove legitimate variants: a registry can present a representative skill, link related alternatives, and preserve attribution where provenance is known. The completeness signals used in RQ3 can help rank these alternatives, while the clone graph reveals relationships that exact hashing misses.}

\revision{For security teams, the 6,376-skill candidate set spanning 2,301 authors illustrates how clone analysis can complement per-skill scanning. Once a scanner flags an artifact, the graph can identify related skills across authors and registries for focused review. Pattern or payload inspection must then determine which members share the relevant security property. This separation preserves the value of clone-based triage without treating similarity as evidence of vulnerability or malware.}

\subsection{\revision{Limitations and Threats to Validity}}

\revision{\noindent\textbf{Evaluation Scope.} Detection is evaluated primarily on the 300-pair \bench{}, with the 450-pair expansion as a scale-robustness check. The method remains ahead of all baselines on the harder set, although its F1 decreases by 5.3 percentage points; unrepresented transformations may yield different performance. The baselines operate on complete skill artifacts and do not cover neural detectors for homogeneous source code. Our model also analyzes only the focal \texttt{SKILL.md}, excluding linked files, screenshots, and commit history.}

\revision{\noindent\textbf{Internal Validity.} The benchmark combines deterministic mutations with fork-mined pairs having verified GitHub provenance (Section~\ref{sec:benchmark}). The stable ranking on the 450-pair expansion and the feature ablation support the main result, but neither source captures every transformation or the full diversity of human edits.}

\revision{\noindent\textbf{External Validity.} RQ3 and RQ4 cover the complete 137,470-skill analysis corpus, but content was unavailable for 21\% of the 196,134 listings (Section~\ref{sec:corpus}) and may differ from the analyzed corpus. Precision may also differ in a deployment corpus dominated by true negatives. Clone-involvement and inflation are threshold-dependent; we report them at the conservative setting in Section~\ref{sec:ecosystem} and retain the threshold-independent 14.6\% exact-duplicate rate as a lower-bound anchor. Prevalence and author concentration may change in future snapshots.}

\revision{\noindent\textbf{Construct Validity.} The source-code taxonomy may not capture every multi-modal similarity pattern. We therefore treat types as interpretive labels, while the quantitative claims rely on binary detection (Section~\ref{subsec:classification}). Clone prevalence does not establish causal reuse, and a pattern match does not establish vulnerability or malware status without provenance or payload-level validation. Finally, ``superseded'' measures relative completeness through content length, code-block count, and NL coverage, not correctness or overall quality, and should support review rather than automatic replacement.}

\section{Related Work}
\label{sec:related_work}

\subsection{Code Clone Detection}

\revision{Code clone detection has been studied extensively for over two decades~\cite{roysurvey}. Representative classical approaches include token-based systems such as CCFinder and CP-Miner~\cite{kamiyaccfinder,licpminer}, text- and tree-based systems such as NiCad and Deckard~\cite{roynicad,jiangdeckard}, and scalable index-based systems such as SourcererCC~\cite{sajnanisourcerercc}.} BigCloneBench~\cite{svajlenkobigclonebench} standardized evaluation, while newer neural approaches such as DeepSim, CCLearner, SAGA, and MAGNET target harder semantic clones~\cite{zhaodeepsim,lilscdc,wusaga,magnetclone}. \revision{Unlike prior work on homogeneous code, \tool{} models YAML, prose, and code jointly to detect channel-specific reuse and build ecosystem-scale clone graphs.}

\subsection{Software Ecosystem Analysis}

\revision{Large-scale ecosystem studies have characterized dependency structure and systemic risk in package ecosystems such as Maven, npm, and PyPI~\cite{zimmermannsmall,witternnpm,kikasdependency,decanempirical,bommaritoempirical,hu2024empirical,zhang2023mitigating,zhang2025fixing}.} Recent work has also begun to characterize the architecture and growth of agent-skill ecosystems~\cite{skillsurvey,skilldataanalysis}. However, prior studies do not model inter-skill clone relationships. \revision{To our knowledge, \tool{} addresses this gap with the first clone graph for agent skills.}

\subsection{Supply Chain Security}

Software supply chain research has shown that vulnerabilities and malicious packages propagate through ecosystem structure rather than through isolated artifacts alone~\cite{liudemystifying,ohmbackstabber,duanmeasuring,zhang2023compatible,dann2023upcy}. In agent ecosystems, recent tools focus on per-skill analysis for vulnerability detection, formal verification, malware identification, prompt injection, and MCP-server risks~\cite{skillscan,skillfortify,maltool,mcpattack,mcpsecurity,skillject,skillinject,skillpromptinjection,promptinjectionagents}. The January 2026 ClawHavoc campaign~\cite{clawhavoc} underscores the practical importance of these risks. \revision{\tool{} complements this line of work with clone-graph analysis: after a security-relevant code pattern is identified in one skill, the clone graph can reveal related skills that may warrant further inspection.}

\section{Conclusion}
\label{sec:conclusion}

\revision{We presented \tool{}, to our knowledge the first multi-modal clone detector for
agent skills, fusing YAML, NL, and code similarity via logistic regression.
\tool{} achieves F1\,=\,0.939 on \bench{} (300 pairs), with 4.2$\times$
higher Type-4 recall than MinHash. Applied to the full corpus of 137,470 skills, it uncovers 1.06 million clone pairs (66.8\% of skills,
95\% cross-author) and yields a conservative 2.45$\times$ inflation estimate, with 67\% of
clone-family skills superseded by a more complete variant. Tracing 938
security-relevant seeds through the clone graph reaches 6,376 related skills
across 2,301 authors, most held by authors other than the seed's own. This
suggests that clone-graph visibility is a promising complement to per-skill
security scanning at ecosystem scale.}

\begin{acks}
This research is supported by the National Research Foundation Singapore, Prime Minister's Office, Singapore, and the Cyber Security Agency under the National Cybersecurity R\&D Programme (NCRP25-P04-TAICeN) and its Campus for Research Excellence and Technological Enterprise (CREATE) programme.
\end{acks}

\section*{Data Availability Statement}
Our replication package is archived on Zenodo:
\par\noindent\url{https://doi.org/10.5281/zenodo.19245910}.

\balance
\bibliographystyle{ACM-Reference-Format}
\bibliography{references}

\end{document}